\documentclass{article}
\usepackage{pdfpages}
\usepackage{xcolor}
\usepackage[mathlines]{lineno}
\usepackage{graphicx}
\usepackage{cite}
\usepackage{caption}
\usepackage{subcaption}
\usepackage{authblk}
\usepackage{amsmath}
\usepackage[top=5cm, bottom=5cm, left=3cm, right=2cm, headsep=1.5cm, heightrounded=true]{geometry}
\usepackage{fancyhdr}
\graphicspath{ {{images/}} }

\title{Barrow Entropy and AdS Black Holes in RPS Thermodynamics   }
	\author[1,2]{Yahya Ladghami\thanks{ \texttt{yahya.ladghami@ump.ac.ma}}}
	\author[1,2]{Brahim Asfour \thanks{\texttt{brahim.asfour@ump.ac.ma}}}
    \author[1,2]{Amine Bouali \thanks{\texttt{a1.bouali@ump.ac.ma }}}
	\author[1,2]{Ahmed Errahmani \thanks{\texttt{ahmederrahmani1@yahoo.fr}}}
	\author[1,2]{Taoufik Ouali\thanks{ \texttt{t.ouali@ump.ac.ma}}}
\affil [1] {Laboratory of Physics of Matter and Radiation, Mohammed I University, BP 717, Oujda, Morocco}
\affil[2]{Astrophysical and Cosmological Center,  BP 717, Oujda, Morocco}
\begin{document}
\maketitle
\begin{abstract}
In this paper, we examine the restricted phase space (RPS) thermodynamics   for charged AdS black holes by considering  the impact of quantum gravity  on the event horizon area. The primary aim of this work is to elucidate the influence of quantum gravitational effects on thermodynamic behaviors, critical phenomena, phase transitions, and the stability of black holes. 
We observe that charged AdS black holes exhibit thermodynamic behavior similar to that of Van der Waals fluids when influenced by quantum gravity. Furthermore, we introduce a novel black hole thermodynamic phenomenon, which we term  ``resistance of phase transitions". Our study uncovers a violation of the homogeneity property of the Smarr relation in RPS thermodynamics due to the effects of quantum gravity.
	
\end{abstract}
\section{Introduction}

Black hole thermodynamics is crucial for comprehending the nature of quantum gravity. The exploration of black hole thermodynamics has garnered  substantial interest, commencing with the works of Hawking, Bekenstein, Bardeen and Carter, who formulated four laws for black holes, analogous to those  of classical thermodynamics  \cite{i1}. They found that  black hole emits a radiation,  has a temperature, and possesses entropy. Moreover, the entropy was found to be proportional to the event horizon area of the black hole \cite{i2,i3,i4}. In addition to explaining the mass of the black hole as internal energy, a fundamental phenomenon has been discovered for asymptotically anti-de Sitter (AdS) black holes: the Hawking-Page phase transition \cite{i5}. This transition occurs between the Schwarzschild-AdS black hole and  pure thermal AdS space. In addition to studying the evolution of black hole shadows \cite{tz1} and the thermodynamics with negative entropy of the black holes \cite{tz2}. Later, this formalism was referred to as ``the traditional black hole thermodynamics" (TBHT). 
\\

The inclusion of the negative cosmological constant as a thermodynamic variable  was a fundamental step in constructing a new formalism for black hole thermodynamics. This formalism, known as ``extended phase space thermodynamics"  (EPST) or black hole chemistry, was introduced by Kastor, Ray, and Traschen \cite{i6}. The cosmological constant is related to the pressure through the equation $P = -\Lambda/8\pi G$. Pressure and its conjugate variable, the thermodynamic volume, together form a new pair of thermodynamic variables, $(P, V)$. In this formalism, the black hole mass is reinterpreted as enthalpy. The EPST formalism plays a important role in  black holes thermodynamics because it reveals various phenomena related to the thermodynamic behavior of black holes. Among these phenomena are phase transitions between  small and large black holes \cite{i7}, multiple critical points \cite{i8,i9}, polymer-type phase transitions \cite{i10}, Joule–Thomson expansion \cite{i11} and stretched quintessence phase\cite{i12}.
\\

Recently, Visser has developed a new version of extended phase space thermodynamics \cite{visser}, within  the framework of Anti-de Sitter/Conformal Field Theory (AdS/CFT) correspondence \cite{AdS/CFT}.  Based on Visser’s thermodynamics, which introduced  the number of colors and the chemical potential  as new variables, Gao and Zhao \cite{RPST 1} proposed a new formalism for black holes thermodynamics by considering the cosmological constant as fixed while the Newton’s constant is treated  as a thermodynamic variable.  This formalism, known as ``restricted phase space (RPS) thermodynamics", is free of a pair pressure-volume $(P,\,V)$, and recovers the interpretation of mass as internal energy. Recently, several studies have utilized this formalism to investigate black holes, including Kerr-AdS black holes \cite{RPST 2}, Taub-NUT AdS black holes and the weak gravity conjecture \cite{RPST 5}, Kerr-Sen-AdS black holes \cite{RPST 3}, 4D-EGB black holes \cite{RPST 4}, black holes  in higher dimensions and higher curvature gravities \cite{RPST 6}, and charged AdS black holes in conformal gravity \cite{RPST 7}.
\\

 In  the context of quantum gravity and black hole physics, Barrow made contributions by incorporating the effects of quantum gravity on the structure of black holes, as detailed in his work \cite{B}. The influence of quantum gravity leads to a deformation in the event horizon area of black holes.	Barrow proposes that, under such effects,  the event horizon area undergoes changes that can be modeled by  a fractal structure. By incorporating fractals into his statement, Barrow modifies the conventional understanding of a smooth and simple event horizon when quantum gravity effects are taken into account.   These changes in the event horizon area are what led Barrow to modify the expression of the entropy of black holes. The effects of quantum gravity, characterizing the deformation  of the event horizon area of black holes, are quantified by a fractal parameter, $\delta$.  Barrow entropy has been extensively discussed in both the context of black holes and cosmology. Among these, we cite constraints imposed by the generalized second law of thermodynamics \cite{B2},  Barrow holographic dark energy  \cite{B3,B4,B5,B6,B7} and  applications in the conjecture of ``gravity-thermodynamics" \cite{GT,GT2}. In addition, it has been used in  the traditional thermodynamics of black holes \cite{BB}, thermodynamic geometries of BTZ black holes and charged AdS black hole with a global mono pole \cite{BB1}, the black holes quasinormal modes and fractal structure \cite{BB2}, Barrow entropy and  the equipartition theorem \cite{BB3}.
 \\
 
Our aim is to explore the evaluation of validity of the RPS formalism for black holes by considering the effect of quantum gravity. Moreover, this paper seeks to examine how quantum gravity influences the thermodynamic properties of charged AdS black holes, while also establishing connections between the fractal parameter, phase transitions, and the stability of these black holes. 
\\
\\

This paper is structured as follows: In Section \ref{ch1}, we introduce restricted phase space thermodynamics incorporating a fractal structure. In Section \ref{ch2}, we present charged AdS black holes with a fractal structure.  Section \ref{ch3} is dedicated to defining the thermodynamic variables and equations of state in RPS thermodynamics for charged AdS black holes. Section \ref{ch4}  explores the impact  of the fractal structure on critical phenomena, phase transitions, and  stability of black holes.   Section \ref{ch5}, presents and discusses our most significant findings related to the current study. In this paper, we adopt the unit  $\hbar = c = k_B = 1$.

\section{RPS Thermodynamics   }
\label{ch1}
In this section, we develop restricted phase space thermodynamics with respect to the Barrow statement. To begin, we establish the formalism of RPS, which was originally formulated by Gao and Zhao \cite{RPST 1}. This formalism is grounded in Visser's thermodynamics \cite{visser} within the framework of AdS/CFT correspondence \cite{AdS/CFT}. Notably, it takes into consideration a fixed cosmological constant and a variable Newton constant, wherein the black hole mass is interpreted as  the internal energy.
\\

The AdS/CFT correspondence states that the partition function of AdS spacetime, $Z_{AdS}$, is equal the CFT dual one, $Z_{CFT}$, \cite{w}. The partition function of CFT is related to the free energy, $W$, by 
\begin{equation}
	\label{c1}
	W= -T ln Z_{CFT}  ,
\end{equation}
where $T$ represents the temperature. Furthermore, we can express the free energy in terms of the central charge $C$ and the chemical potential, $\mu$, as follows \cite{visser}
\begin{equation}
	\label{c2}
	W= \mu\, C.
\end{equation}
The central charge of CFT  is defined as \cite{RPST 1, RPST 2} 
\begin{equation}
	\label{AZ}
C= \frac{l^{d-2}}{G},
\end{equation}
where $l$ is the AdS radius, $d$ is the dimension of the bulk spacetime, and $G$ is the Newton constant. The partition function of AdS spacetime is related to the Euclidean action $I_E$ \cite{IE,Rn} by
\begin{equation}
	\label{c3}
	I_E= - ln Z_{AdS}. 
\end{equation}
For  charged and  rotating AdS black holes in a $d$ dimension, the Euclidean action writes 
\begin{equation}
	\label{c4}
	I_{\mathrm{E}}=-\frac{1}{16 \pi G} \int d^{d} x \sqrt{-g}\left(R-2 \Lambda-F_{\mu \nu} F^{\mu \nu}\right),
\end{equation}
 where $R$ is the  Ricci scalar, $\Lambda$ is the cosmological constant, and $F_{\mu \nu}$ represents the Maxwell strength field. From the formulation of the AdS/CFT correspondence, as expressed in Eqs. \eqref{c1}-\eqref{c3} with the equality $Z_{AdS}=Z_{CFT}$, one can infer that the free energy is correlated with the parameters characterizing black holes by \cite{Rn, Ke}
 \begin{equation}
 	\label{c4}
 \mu C=TI_E =M-T S-\Phi Q-\Omega J,
 \end{equation}
where $M$ denotes the black hole mass and $S$ corresponds to the Bekenstein-Hawking entropy.  Additionally,  $\Phi$ is the electric potential in the event horizon  due to the black hole's electric charge, denoted by Q.  $\Omega$ and $J$ represent  the angular velocity and angular momentum, respectively. The Smarr relation within the RPS formalism can be written as follows
 \begin{equation}
 	\label{first}
 	M=T S+\tilde{\Phi} \tilde{Q}+\Omega J + \mu C.
 \end{equation}
  where we have introduced the rescaled electric potential, $\tilde{\Phi}$, and the rescaled  electric charge,  $\tilde{Q}$, as 

\begin{equation}
	\tilde{\Phi}=\frac{\Phi }{\sqrt{C}} \quad \text{and} \quad \tilde{Q}=Q\,\sqrt{C},
\end{equation}
respectively. The first law in the RPS formalism is expressed as follows \cite{RPST 1}
\begin{equation}
	\mathrm{d} M=T\mathrm{~d} S+\Omega \mathrm{d} J+\tilde{\Phi} \mathrm{d} \tilde{Q}+\mu \mathrm{d} C.
\end{equation}
After obtaining the first law and the Smarr relation, we will reformulate the restricted phase space thermodynamics by incorporating Barrow's corrections. Barrow has demonstrated that the effects of quantum gravity can induce deformations on the black hole's surface, which are described by a fractal structure. This, in turn, leads to a modification in the expression of the black hole entropy \cite{B}, dubbed   Barrow entropy, as
\begin{equation}
		\label{SB}
	S_B=\left(\frac{A}{A_{\text{PI}}}\right)^{1+\frac{\delta}{2}},
\end{equation}
where $A$ represents the event horizon area of a black hole, $A_{\text{PI}} = 4G$ is the Planck area, and $\delta$ refers to the parameter quantifying  deformations of quantum gravity on the event horizon area, with $0 \leq \delta \leq 1$. When $\delta = 0$, the Bekenstein-Hawking entropy is restored, signifying the absence of a fractal structure. In the case where $\delta = 0$, we recover the normal study of Reissner-Nordström AdS black holes, i.e., without a fractal structure of the event horizon area\cite{RPST 1}. Conversely, when $\delta = 1$, it signifies the most deformed and complex  fractal structure of the black hole's event horizon area.
\\

In black hole thermodynamics, temperature is the conjugate quantity of entropy. Consequently, the Barrow temperature $T_B$ is defined as 
\begin{equation}
	\label{TB}
	T_B = \frac{\partial M}{\partial S_B}.
\end{equation}
We establish the relationship between the modified quantities $\left( S_B, T_B\right) $ and the  ordinary quantities $\left( S, T\right) $, using  Eqs. \eqref{SB} and \eqref{TB}, as 
\begin{equation}
	\label{TS1}
	T\,S=\left(1 + \frac{\delta }{2}\right) T_B\,S_B.
\end{equation}
We substitute Eq. \eqref{TS1}  into Eq. \eqref{first}, and thereby we derive the Smarr relation in terms of the modified  quantities
\begin{equation}
	\label{f1}
	M=\left(1 + \frac{\delta }{2}\right)T_B\, S_B+\tilde{\Phi} \tilde{Q}+\Omega J + \mu\, C.
\end{equation}
In this Smarr relation, M is a homogeneous function of $(T_B,S_B)$ with an order of $\left(1 + \delta/2\right)$, and of $(\tilde{\Phi},\, \tilde{Q})$, $(\mu,\,C)$, and $(\Omega,\, J)$ with an order of 1. This result contradicts the nature of the Smarr relation in the RPS formalism in normal cases \cite{RPST 1,RPST 2,RPST 5, RPST 3,RPST 4},  where $M$ is a homogeneous function of order 1 of both quantities. The homogeneity property of the Smarr relation in RPS thermodynamics is violated by the effects of quantum gravity, represented by the fractal parameter. Mathematically, these effects   are determined by the addition of the element $(T_B,S_B)$ with the factor, $\delta/2$, signifying a departure from the homogeneity of the Smarr relation in the RPS formalism. For $\delta = 0$, we recover the Smarr relation of  the ordinary RPS formalism \cite{RPST 1}. We can construct the first law for charged AdS black holes, where $J=0$, by differentiating the mass function, where $M=M(S_B, \tilde{Q},C)$.
\begin{equation}
	\label{FL}
	d M=T_B d S_B+\tilde{\Phi} d \tilde{Q}+ \mu d C.
\end{equation}

\section{Charged AdS Black Holes}
 \label{ch2}
In this section, we study the effects of the fractal structure on  charged black holes in the Anti-de Sitter spacetime in the context of restricted phase space thermodynamics. The action describing  Einstein-Maxwell theory in 4-dimensional  AdS spacetime, is given by
\begin{equation}
	\label{c4}
	I=-\frac{1}{16 \pi G} \int d^{4} x \sqrt{-g}\left(R-2 \Lambda-F_{\mu \nu} F^{\mu \nu}\right),
\end{equation} 
where R is the Ricci scalar, $\Lambda = -3/l^2$ is the cosmological constant, and $F^{\mu \nu} = \partial^\mu A^\nu - \partial^\nu A^\mu$ is the Maxwell field,
with the electromagnetic potential given by
\begin{equation}
	A^\mu=(\frac{Q}{r}, 0,0,0).
\end{equation}
The metric for  a charged AdS black holes is given by
\begin{equation}
	d s^2=-f(r) d t^2+\frac{1}{f(r)} d r^2+r^2\left(d \theta^2+\sin ^2 \theta d \phi^2\right),
\end{equation}
where the metric function is defined as
\begin{equation}
	\label{function}
	f(r)=1-\frac{2 G M}{r}+\frac{G Q^2}{r^2}+\frac{r^2}{l^2}.
\end{equation}
These black holes under consideration possess two horizons, denoted as $r_\pm$. The event horizon, $r_+$,  represents the exterior radius, while $r_-$ represents the interior horizon. The expression for the black hole's mass, obtained by solving the equation $f(r_+)=0$, writes 
\begin{equation}
	\label{Mass}
	M=\frac{r_{+}}{2 G}\left(1+\frac{G Q^2}{r_{+}^2}+\frac{r_{+}^2}{l^2}\right).
\end{equation}
The Hawking temperature of the charged black hole and the Bekenstein-Hawking entropy are given by
\begin{equation}
	\label{T}
	T=\frac{f^{\prime}\left(r_{+}\right)}{4 \pi}=\frac{r_{+}^2+3 r_{+}^4 l^{-2}-G Q^2}{4 \pi r_{+}^3},
\end{equation}
\begin{equation}
	\label{AZ1}
	S= \frac{\pi r_+^2}{G},
\end{equation}
respectively. Due to the quantum gravity effects on the event horizon area, the entropy of the black hole becomes the Barrow entropy
\begin{equation}
	\label{Sb}
	S_B= \left( \frac{\pi r_+^2}{G}\right)^{1+\frac{\delta}{2}},
\end{equation}
and the Barrow temperature, from Eq. \eqref{TB}, is expressed as
\begin{equation}
	\label{Tb}
	T_B = \frac{r_{+}^2+3 r_{+}^4 l^{-2}-G Q^2}{2 \pi r_{+}^3\left(2+ \delta \right)\left( \frac{\pi r_+^2}{G}\right)^{\frac{\delta}{2}} }.
\end{equation}
For $\delta=0$, we  recover the Hawking temperature and the Bekenstein-Hawking entropy. However, in the case of the most deformed fractal structure, i.e. $\delta=1$, we obtain $S_B \propto r_+^3$. This is similar to the entropy of  black hole in $(4+1)$-dimensions \cite{Sr}. This result, due to a maximum fractal structure, breaks the holographic principle which states that the entropy of a black hole is proportional to its surface. From Eq. \eqref{f1}, we obtain the Smarr relation of a charged AdS black hole with Barrow correction in the RPS thermodynamics as
\begin{equation}
	\label{f2}
	M=\left(1 + \frac{\delta }{2}\right)T_B\, S_B+\tilde{\Phi} \tilde{Q}+   \mu C,
\end{equation} 
and the first law is 
\begin{equation}
	\label{Fl1}
	d M=T_B d S_B+\tilde{\Phi} d \tilde{Q} + \mu d C.
\end{equation}
Hereafter, we will denote the Barrow quantities $(S_B,\,T_B,)$ by $(S,\,T)$, respectively.
\section{Equations of State}
\label{ch3}
In this section, we will delve further into the study of charged AdS black holes, incorporating  influences of quantum gravity through the framework of RPS thermodynamics. We will establish thermodynamic parameters and formulate the equations of state.\\
Using Eq. \eqref{Sb},  the black hole mass, Eq \eqref{Mass},  in terms of extensive variables  and the fractal parameter, $\delta $, becomes

\begin{equation}
	\label{MassB}
	M\left( S,\,C,\,\tilde{Q},\,\delta\right) =\frac{\pi  C S^{\frac{2}{\delta +2}}+\pi ^2 \tilde{Q}^2+S^{\frac{4}{\delta +2}}}{2 \pi ^{3/2} l \sqrt{C S^{\frac{2}{\delta +2}}}}.
\end{equation} 
Through the first law, \eqref{Fl1}, all thermodynamic quantities can be obtained as follows
	\begin{equation}
		T=\left(\frac{\partial M}{\partial S}\right)_{C, \tilde{Q}, \delta}, \qquad 	\tilde{\Phi}=\left(\frac{\partial M}{\partial \tilde{Q}}\right)_{S, C, \delta}, \qquad	\mu=\left(\frac{\partial M}{\partial C}\right)_{S, \tilde{Q}, \delta}.
	\end{equation}
Finally, we determine the equations of state describing charged AdS black holes in the presence of quantum gravity effects 
 \begin{equation}
 	\label{5}
 		T=\frac{\pi  C S^{\frac{2}{\delta +2}}-\pi ^2 \tilde{Q}^2+3 S^{\frac{4}{\delta +2}}}{2 \pi ^{3/2} (\delta +2) l S \sqrt{C S^{\frac{2}{\delta +2}}}},
 \end{equation}
\begin{equation}
	\label{5io}
	\tilde{\Phi}=\frac{\sqrt{\pi } \tilde{Q}}{l \sqrt{C S^{\frac{2}{\delta +2}}}},
\end{equation}
\begin{equation}
	\label{6y}
	\mu=-\frac{-\pi  C S^{\frac{2}{\delta +2}}+\pi ^2 \tilde{Q}^2+S^{\frac{4}{\delta +2}}}{4 \pi ^{3/2} C l \sqrt{C S^{\frac{2}{\delta +2}}}},
\end{equation}
From Eq. \eqref{MassB} and Eqs. \eqref{5}-\eqref{6y}, it is evident that $M$, $T$, $\tilde{\Phi}$, and $\mu$ depend on the extensive variables of $S$, $\tilde{Q}$, and $C$. When the variables $S$, $\tilde{Q}$, and $C$ undergo rescaling, namely $S \rightarrow \lambda S$, $\tilde{Q} \rightarrow \lambda \tilde{Q}$, $C \rightarrow \lambda C$,   $M$ looses its first-order homogeneity. Furthermore, $T$, $\tilde{\Phi}$, and $\mu$ no longer exhibit zero-order homogeneity, contradicting the findings in \cite{RPST 1}.
In the absence of a fractal structure, $\delta = 0$, we recover the equation of state for the standard  RPS formalism \cite{RPST 1}. Conversely, when the structure of black hole horizons  exhibit its maximum fractal structure, $\delta = 1$, we obtain
\begin{equation}
	\label{T11}
	T= \frac{\pi  C S^{2/3}-\pi ^2 \tilde{Q}^2+3 S^{4/3}}{6 \pi ^{3/2} l S \sqrt{C S^{2/3}}},
\end{equation}
\begin{equation}
	\tilde{\Phi}= \frac{\sqrt{\pi } \tilde{Q}}{l \sqrt{C S^{2/3}}},
\end{equation}
\begin{equation}
	\mu=-\frac{-\pi  C S^{2/3}+\pi ^2 \tilde{Q}^2+S^{4/3}}{4 \pi ^{3/2} C l \sqrt{C S^{2/3}}}.
\end{equation}
Through the equations of state, Eqs. \eqref{5}-\eqref{6y},  the effect of the fractal structure parameter appears clear, as it changes the degrees of extensive variables, $(S,\, \tilde{Q},\, C)$.
\section{Thermodynamic Processes and Phase Transitions}
\label{ch4}
In this section, we will investigate the impact of the  fractal structure  on thermodynamic processes, critical phenomena, and the stability of black holes. Our approach involves determining the critical parameters by solving the following system,
\begin{equation}
	\label{6}
	\left(\frac{\partial T}{\partial S} \right)_{C, \tilde{Q}, \delta}=0, \qquad 	\left(\frac{\partial^2 T}{\partial S^2} \right)_{C, \tilde{Q}, \delta}=0.
\end{equation}
From Eqs.\eqref{5} and \eqref{6}, we find, for $\delta\neq 1$, the critical entropy $S_c$ 
\begin{equation}
	\label{ya1}
	S_c= \left(\frac{\pi\,  C\, (\delta +1)}{6 (1-\delta )}\right)^{\frac{\delta }{2}+1},
\end{equation}
and the critical re-scaled electric charge $\tilde{Q}_c$
\begin{equation}
		\label{ya2}
	\tilde{Q}_c= \frac{C (\delta +1)}{ 2 \sqrt{\left(3-3 \delta\right)  \left(  \delta +3\right) }}.
\end{equation}
Hence, at critical points, the critical temperature $T_c$ writes
\begin{equation}
		\label{ya3}
	T_c=\frac{2^{\frac{\delta +3}{2}} \sqrt{3 (\delta +1)}}{l\, (\delta +1) (\delta +2) (\delta +3) \pi ^{\frac{\delta }{2}+1} \sqrt{1-\delta }  \left(\frac{C (\delta +1)}{3-3 \delta }\right)^{\delta /2}}.
\end{equation} 
 When considering the maximum fractal level ($\delta = 1$), the critical parameters diverge, indicating the absence of any  critical phenomena. Furthermore, we can determine the entropy at which first-order phase transitions occur by solving the following equation
 \begin{equation}
 	\label{6A}
 	\left(\frac{\partial T}{\partial S} \right)_{C, \tilde{Q}, \delta}=0,
 \end{equation}
after excluding the imaginary solutions, we find two entropy values
  \begin{equation}
  S_1=\left(\frac{\pi}{6}\right)^{1 + \frac{\delta}{2}} \left(\frac{C + C\delta + \sqrt{C^2 (1 + \delta)^2 + 12 \tilde{Q}^2 (-1 + \delta) (3 + \delta)}}{1 - \delta}\right)^{{1 + \frac{\delta}{2}}}
    \end{equation}
and 
 \begin{equation}
S_2=\left(\frac{\pi}{6}\right)^{1 + \frac{\delta}{2}} \left(\frac{C + C\delta - \sqrt{C^2 (1 + \delta)^2 + 12 \tilde{Q}^2 (-1 + \delta) (3 + \delta)}}{1 - \delta}\right)^{{1 + \frac{\delta}{2}}}
\end{equation}
\begin{figure}[htp]
	\centering
	\includegraphics[scale=0.6]{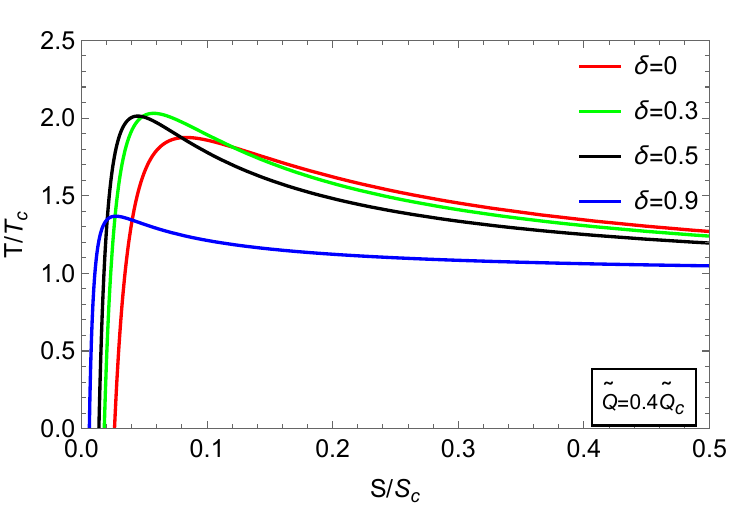}
	\includegraphics[scale=0.6]{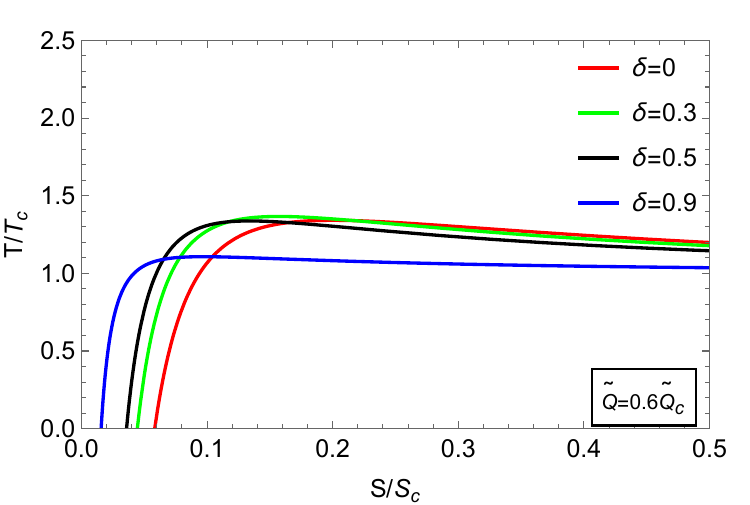}
	\includegraphics[scale=0.6]{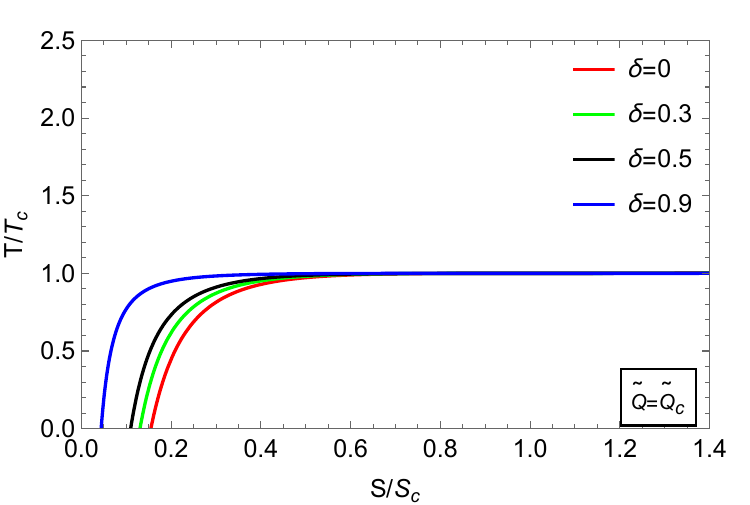}
		\includegraphics[scale=0.6]{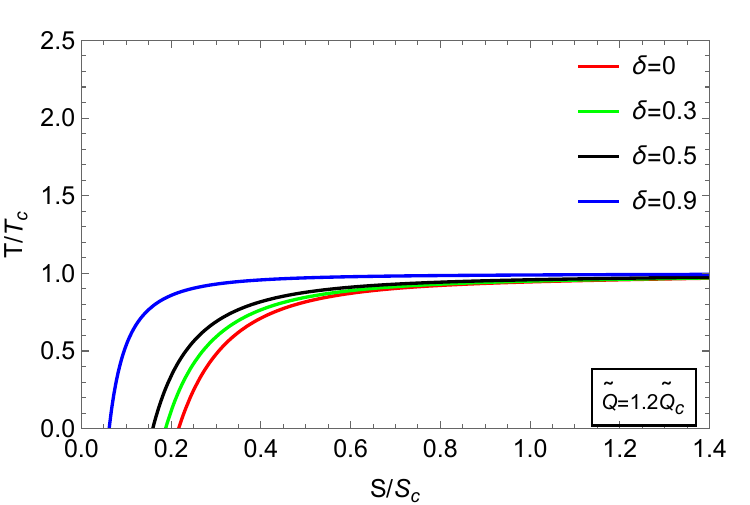}
	\caption{$T-S$ curves for different values of $\delta$ and $\tilde{Q}$.}
	\label{TS}
\end{figure}
\\

In Fig. \ref{TS}, where we plot the normalized Barrow temperature $T/T_c$ as a function of the normalized Barrow entropy $S/S_c$, three distinct types of $T$-$S$ curves can be identified. The first type, characterized by $\tilde{Q} < \tilde{Q}_c$, corresponds to a first-order phase transition between small and large black holes while passing through an intermediate black hole. The second type, occurs when  the electric charge of the black hole  reaches  its critical value,  indicating a second-order phase transition from small black holes  to  large black holes. For the last type, where  $\tilde{Q} > \tilde{Q}_c$,  there are no phase transitions.  We concluded that, within the RPS formalism with the effect of quantum gravity on the event horizon area, the thermodynamics of charged AdS black holes exhibit similarities to the thermodynamics of a Van der Waals fluid. The value of the central charge does not affect the behavior of thermodynamics and the temperature evolution. This property is similar to ``the law of corresponding states" in thermodynamics for ordinary matter, where the thermodynamic behavior is completely independent of the number of particles.
\begin{figure}[htp]
	\centering
	\includegraphics[scale=0.6]{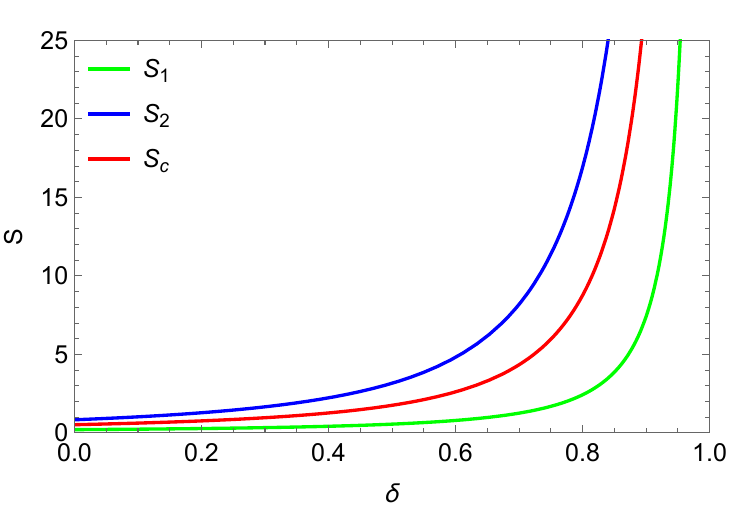}
	\caption{Critical entropy and phase transition entropy as a functions of the fractal parameter,  with $C=1$}
	\label{rph}
\end{figure}
\\

In Fig. \ref{rph}, we plot the evolution of the critical entropy $S_c$ characterizing second-order phase transitions and the entropies of first-order phase transitions $S_1$ and $S_2$ in terms of the fractal parameter.We observe that as the value of $\delta$ increases, both the critical entropy $S_c$ and the entropies of phase transitions also increase. When the deformation reaches its maximum value $\delta=1$, the critical quantities $(S_c, \tilde{Q}_c, T_c)$ and the entropies of phase transitions $(S_1, S_2)$ diverge, indicating that the phase transitions do not occur under these circumstances. The deformation caused by quantum gravity delays phase transitions, critical phenomena, and even prevents them from occurring when the deformation reaches its maximum.Thus, this deformation due to quantum gravity effects on the event horizon behaves as a ``phase transitions resistance''.
\\

Furthermore, the nature of phase transitions is described by studying the free energy. We express  the Barrow free energy, denoted as $F$, as follows 
 \begin{equation}
 	\label{F74}
 	F\left( S,\,C,\,\tilde{Q},\,\delta\right) = 	M\left( S,\,C,\,\tilde{Q},\,\delta\right)  - TS.
 \end{equation}
Next, we intend to  plot the normalized Barrow free energy $F/F_c$ against  the normalized Barrow temperature $T/T_c$, where $F_c$ is the value of the free energy that corresponds to the critical quantities.
\begin{figure}[htp]
	\centering
	\includegraphics[scale=0.6]{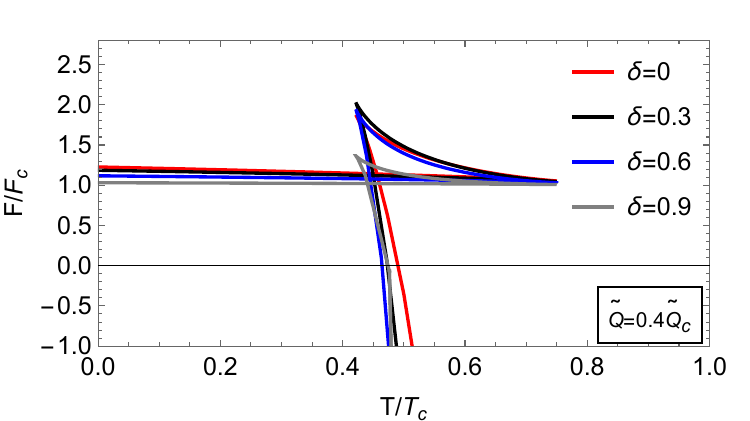}
	\includegraphics[scale=0.6]{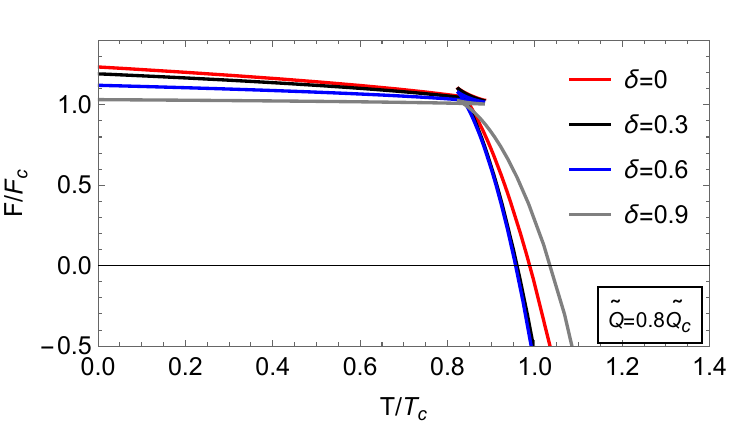}
	\includegraphics[scale=0.6]{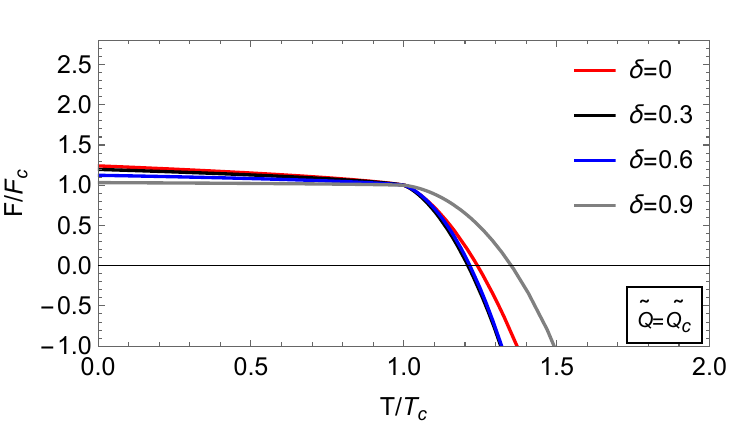}
	\includegraphics[scale=0.6]{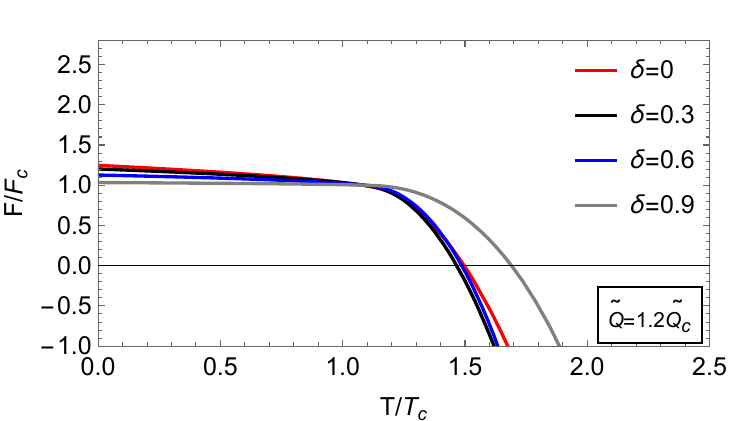}
	\caption{$F-T$ curves for different values of $\delta$ and $\tilde{Q}$.}
	\label{FS}
\end{figure}
\\

Fig. \ref{FS} illustrates the evolution of  Barrow free energy, $F$, as a function of the Barrow temperature for different values of the electric charge and the fractal parameter. When $\tilde{Q} < \tilde{Q}_c$, a swallowtail pattern is observed, indicating a first-order phase transition. However, as the fractal parameter increases, the area of the swallowtail decreases. However, this pattern does not appear in other cases, $\tilde{Q} \ge \tilde{Q}_c$, regardless of the temperature value.
\\

One key objective of black hole thermodynamics is to investigate the stability of black holes. The stability of black holes is determined by analyzing their heat capacity. A positive heat capacity indicates stability, whereas a negative value implies instability. In this paper, we study the effect of quantum gravity on the thermodynamic behavior of black holes, including their stability. To achieve this, we introduce the Barrow heat capacity, denoted as $\xi$, which incorporates the influences of quantum gravity, as expressed by the following relation.
\begin{equation}
	\label{Heat}
	\xi= T\, \left(\frac{\partial S}{\partial T}\right)_{C, \tilde{Q}, \delta}.
\end{equation}
Through Eqs. \eqref{T11} and \eqref{Heat}, we can plot the evolution of the Barrow heat capacity as a function of the entropy  in various scenarios involving electric charge and  fractal parameter.
\begin{figure}[htp]
	\centering
	\includegraphics[scale=0.6]{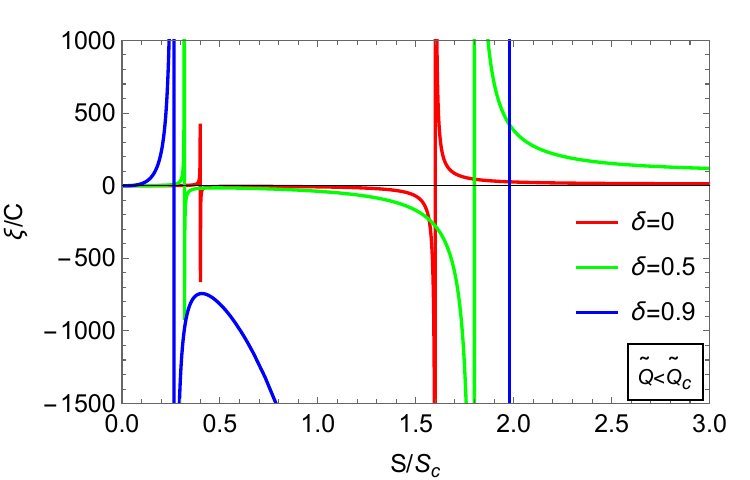}
	\includegraphics[scale=0.6]{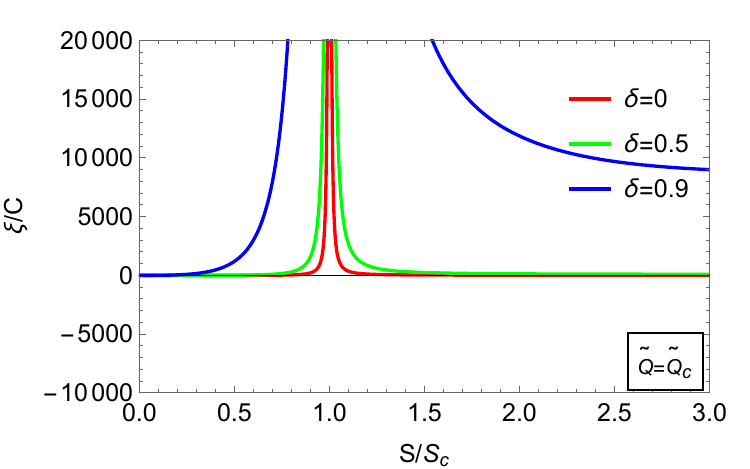}
	\includegraphics[scale=0.6]{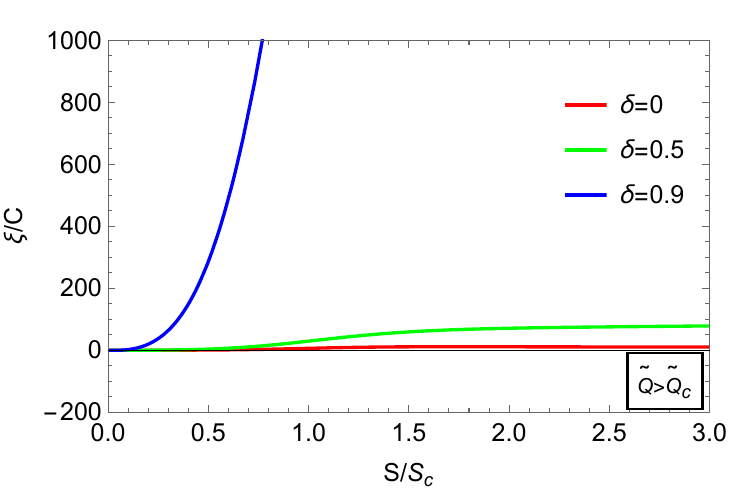}
	\caption{Variation of $\xi/C$ as a function of $S/S_c$ for different values of the fractal parameter and electric charge, with $C=1$}
		\label{CS}
\end{figure}
\\

 Fig. \ref{CS}, represents the evolution of  the heat capacity divided by the central charge as a function of the entropy divided by the critical entropy.  The central charge is always positive, so the sign of  $\xi/C$ determines the sign of the heat capacity. In Fig. \ref{CS}, three sub-figures are presented, each corresponding to a particular case of electric charge. For the first case, i.e.  $\tilde{Q} < \tilde{Q}_c$, indicates a first-order phase transition among Small/Intermediate/Large black holes. We observe that both small and large black holes demonstrate stability, while intermediate black holes exhibits instability. In the second case,  $\tilde{Q} = \tilde{Q}_c$, i.e. a second-order phase transition between Small/Large black holes, we find stability in both small and large black holes is observed. In the case of $\tilde{Q} > \tilde{Q}_c$,  black holes are stable. When considering the impact of a fractal structure on the stability of black holes,  we notice that as $\delta$ increases, the region of intermediate black holes i.e. unstable region also expands,  and the value of the heat capacity increases for $\tilde{Q} \ge \tilde{Q}_c$.
  
 
\section{Discussions and Conclusions}
\label{ch5}
In our analysis, we have reformulated the thermodynamics of RPS, considering the influence of  quantum gravity effects. Barrow \cite{B} concluded that the area of the event horizon is subject to deformations generated by quantum gravity effects and modeled using the fractal structure. This deformation leads to a modification in the expression for the area of the event horizon, consequently reformulating the entropy of black holes as Barrow entropy.
\\

This Study of  charged AdS black holes within the framework of the RPS formalism has allowed us to uncover the influence of quantum gravity and its impact on the thermodynamic properties of black holes. Our study was conducted using the RPS formalism, which is a development version of Visser's thermodynamics with certain restrictions. We considered the AdS radius as a constant while treating Newton's constant as a variable. This thermodynamics is constructed within  the framework of the AdS/CFT correspondence.
\\

Through the  entropy and temperature in fractal structure, we find the Smarr relation and the first law.  In  the RPS thermodynamics with the fractal structure, the black hole mass is interpreted as internal energy. While the Smarr relation does not remain homogeneous of the first order, due to the appearance of the fractal structure parameter in the relation. In addition, if the deformation resulting from the effects of quantum gravity is at its maximum, the Barrow entropy of black holes will be proportional to the cubic size of  black holes as in five dimensions \cite{Sr}. 
\\

We have selected charged AdS black holes as an application and we have derived  thermodynamic quantities and  equations of state in terms of the parameter that characterizes the fractal structure. Additionally, we have investigated critical quantities within this context. After studying the evolution of the temperature of black holes and their heat capacity in terms of the entropy for different values of the electric charge  and the fractal parameter, we have reached the following conclusion: For the case $\tilde{Q}> \tilde{Q}_c$, no phase transition occurs and  black holes are stable. Second-order phase transitions occur between small and large black holes, if the electric charge equals the critical one. For, $\tilde{Q}< \tilde{Q}_c$, first-order phase transitions occur between  small and  large black holes, passing through   intermediate black holes, where  small and  large black holes are stable and the  intermediate black holes are unstable. This behavior of thermodynamics is similar to that of Van der Waals fluids. One important effect of the fractal structure on thermodynamic properties is that at maximum complexity and deformation, the critical quantities diverge. In other words, critical phenomena do not occur. The closer the parameter is to 1, the more delayed the phase transition becomes. Deformations caused by the effect of quantum gravity on the event horizon hinder the phase transition. Therefore, these deformations represent a ``resistance to phase transitions". In the absence of deformations, the situation is associated with Reissner-Nordström AdS black holes  as outlined in \cite{RPST 1}.
\\

In upcoming papers, we will extend our examination of how quantum gravity impacts black holes to encompass various types of black holes and diverse gravity theories.

\end{document}